\documentclass[aps,prd,preprint,tightenlines,superscriptaddress,showpacs]{revtex4}

\usepackage{graphicx} 

\def\myspecial#1{}                   

\renewcommand{\bar}[1]{\overline{#1}{}}

\newcommand{\dm}{\Delta m}
\newcommand{\dt}{\Delta t}

\newcommand{\qq}{q{\bar q}}
\newcommand{\ks}{K_S}

\newcommand{\db}{\bar{D}^0}

\newcommand{\bdbpn}{B^0\to \db h^0}
\newcommand{\kspipi}{\ks\pi^+\pi^-}

\newcommand{\dbkspipi}{\db\to\ks\pi^+\pi^-}
\newcommand{\bdbkspipi}{\bar{B}^0\to \left( \kspipi \right)_D h^0}
\newcommand{\fq}{\ensuremath{q}}

\begin{document}

\preprint{
  \vbox{
    \hbox{   }
    \hbox{hep-ph/0503174}
    \hbox{\today}
  }
}

\title{\quad\\[0.5cm] \boldmath
  A Method to Measure $\phi_1$ Using $\bar{B}^0 \to D h^0$ With Multibody $D$ Decay
}

\author{Alex Bondar\footnote{\tt a.e.bondar@inp.nsk.su}}
\affiliation{Budker Institute of Nuclear Physics, Novosibirsk}
\author{Tim Gershon\footnote{\tt gershon@bmail.kek.jp}}
\affiliation{High Energy Accelerator Research Organization (KEK), Tsukuba}
\affiliation{Department of Physics, University of Tokyo, Tokyo}
\author{Pavel Krokovny\footnote{\tt krokovny@bmail.kek.jp}}
\affiliation{High Energy Accelerator Research Organization (KEK), Tsukuba}

\myspecial{!userdict begin /bop-hook{gsave 280 50 translate 0 rotate
    /Times-Roman findfont 18 scalefont setfont
    0 0 moveto 0.70 setgray
    (\mySpecialText)
    show grestore}def end}

\begin{abstract} 
  \noindent
  We describe a new method to measure the angle $\phi_1$ of the CKM 
  Unitarity Triangle using amplitude analysis of the multibody 
  decay of the neutral $D$ meson produced via $\bar{B} \to D h^0$
  colour-suppressed decays.
  The method employs the interference between $D^0$ and $\bar{D}^0$ 
  to directly extract the value of $2\phi_1$,
  and thus resolve the ambiguity between $2\phi_1$ and $\pi-2\phi_1$
  in the measurement of $\sin(2\phi_1)$ using $\bar{B}^0 \to J/ \psi \ks$. 
  We present a feasibility study of this method using Monte Carlo simulation.
\end{abstract}


\maketitle

{\renewcommand{\thefootnote}{\fnsymbol{footnote}}}
\setcounter{footnote}{0}

\section{Introduction}

Precise determinations of the Cabibbo-Kobayashi-Maskawa (CKM) 
matrix elements~\cite{ckm} are important to 
check the consistency of the Standard Model and search for new physics.
The value of $\sin(2\phi_1)$, 
where $\phi_1$ is one of the angles of the Unitarity Triangle~\cite{pdg_review} 
is now measured with high precision: 
$\sin(2\phi_1) = 0.731 \pm 0.056$~\cite{sin2phi1}. 
However, this measurement contains an intrinsic ambiguity: $2\phi_1 \longleftrightarrow \pi-2\phi_1$.
Various methods to resolve this ambiguity 
have been introduced~\cite{phi_ambig}, 
but they require very large amounts of data 
(some impressive first results notwithstanding~\cite{babar_psikstar}).

We suggest a new technique based on the analysis of $\bar{B}^0 \to D h^0$,
followed by the three-body decay of the neutral $D$ meson, $D \to \ks\pi^+\pi^-$. 
Here we use $h^0$ to denote a light neutral meson, such as $\pi^0, \eta, \rho^0, \omega$.
The modes $\bar{B}^0 \to D_{CP} h^0$,
utilizing the same $B$ decay but requiring the $D$ meson to 
be reconstructed via $CP$ eigenstates,
have previously been proposed as ``gold-plated'' modes to search for 
new physics effects~\cite{grossman_worah}.
Such effects may result in deviations from the Standard Model
prediction that $CP$ violation effects in $b \to c\bar{u}d$ transitions
should be very similar to those observed in $b \to c\bar{c}s$ transitions,
such as $\bar{B}^0 \to J/ \psi \ks$.
Detailed considerations have shown that the contributions
from $b \to u\bar{c}d$ amplitudes,
which are suppressed by a factor of approximately $0.02$~\cite{rd},
can be taken into account~\cite{fleischer}.
Consequently, within the Standard Model,
studies of $\bar{B}^0 \to D_{CP} h^0$ can give a measurement of $\sin(2\phi_1)$ 
that is more theoretically clean than that from $\bar{B}^0 \to J/ \psi \ks$.
However, these measurements still suffer from the ambiguity mentioned above.

In the case that the neutral $D$ meson produced in $\bar{B}^0 \to D h^0$
is reconstructed in a multibody decay mode,
with known decay model,
the interference between the contributing amplitudes 
allows direct sensitivity to the phases.
Thus $2\phi_1$, rather than $\sin(2\phi_1)$ is extracted,
and the ambiguity $2\phi_1 \longleftrightarrow \pi-2\phi_1$ can be resolved.
This method is similar to that used to extract $\phi_3$,
using $B^\pm \to D K^\pm$ followed by multibody $D$ decay~\cite{anton,ggsz};
in the $\phi_3$ analysis the ambiguities in the result are also 
reduced compared to more traditional techniques~\cite{glw,ads}.
In addition, with current $B$ factory statistics, 
better precision on $\phi_3$ is obtained using multibody $D$ decay.

There are a large number of different final states 
to which this method can be applied.
In addition to the possibilities for $h^0$,
and the various different multibody $D$ decays which can be used,
the method can also be applied to $\bar{B}^0 \to D^* h^0$.
In this case, the usual care must be taken to distinguish 
between the decays $D^* \to D\pi^0$ and $D^* \to D\gamma$~\cite{bg}.

In this paper we concentrate primarily on the decay 
$\bar{B}^0 \to D \pi^0$ with $D \to \kspipi$
(and denote the decay chain as $\bar{B}^0 \to \left( \kspipi \right)_D \pi^0$).
This multibody $D$ decay has been shown, in the $\phi_3$ analysis, 
to be particularly suitable for Dalitz plot studies.
In the remainder of the paper, 
we first give an overview of the relevant formalism,
and then turn our attention to Monte Carlo simulation studies of 
$\bar{B}^0 \to \left( \kspipi \right)_Dh^0$.
We attempt to include all experimental effects, 
such as background, resolution, flavour tagging, and so on,
in order to test the feasibility of the method.
Based on these studies, we estimate the precision 
with which $2\phi_1$ can be extracted with the current $B$ factory statistics.

Before turning to the details, we note that this method 
can also be applied to other neutral $B$ meson decays 
with a neutral $D$ meson in the final state.
In particular, the decay $\bar{B}^0 \to D \ks$ has contributions
from $b \to c\bar{u}s$ and $b \to u\bar{c}s$ amplitudes, 
which have a relative weak phase difference of $\phi_3$.
Thus a time-dependent Dalitz plot analysis of 
$\bar{B}^0 \to \left( \kspipi \right)_D \ks$ 
can be used to simultaneously measure $\phi_1$ and $\phi_3$~\cite{ggssz}, 
and can test the Standard Model prediction that
$CP$ violation effects in $b \to c\bar{u}s$ transitions should be, 
to a good approximation, the same as those in $b \to c\bar{c}s$ transitions.
Furthermore, modes such as $B^0_s \to \left( \kspipi \right)_D \phi$ 
can in principle
be used to measure the weak phase in $B^0_s$\textendash$\bar{B}^0_s$ mixing.
However, our feasibility study is not relevant to $B^0_s$ decay modes,
which cannot be studied at a $B$ factory operating at the $\Upsilon(4S)$ resonance.

\section{Description of the method}

Consider a neutral $B$ meson,
which is known to be $\bar{B}^0$ at time $t_{\rm tag}$
(for experiments operating at the $\Upsilon(4S)$ resonance,
such knowledge is provided by tagging the flavour of the other $B$ meson
in the $\Upsilon(4S) \to B\bar{B}$ event).
At another time $t_{\rm sig}$ 
the amplitude content of the $B$ meson is given by~\cite{asy}
\begin{equation}
  \label{eq:b0bar_evo}
  \left| \bar{B}^0(\dt) \right> = 
  e^{-\left|\dt\right|/2\tau_{B^0}}
  \left(
    \left| \bar{B}^0 \right> \cos(\dm\dt/2) - 
    i \frac{p}{q} \left| B^0 \right> \sin(\dm\dt/2)
  \right),
\end{equation}
where $\dt = t_{\rm sig} - t_{\rm tag}$, 
$\tau_{B^0}$ is the average lifetime of the $B^0$ meson,
$\dm$, $p$ and $q$ are parameters of $B^0 - \bar{B}^0$ mixing
($\dm$ gives the frequency of $B^0 - \bar{B}^0$ oscillations,
while the eigenstates of the effective Hamiltonian in the 
$B^0 - \bar{B}^0$ system are 
$\left| B_\pm \right> = p \left| B^0 \right> \pm q \left| \bar{B}^0 \right>$),
and we have assumed $CPT$ invariance and 
neglected terms related to the $B^0 - \bar{B}^0$ lifetime difference~\cite{bs}.
In the following we drop the terms of $e^{-\left|\dt\right|/2\tau_{B^0}}$.

\begin{figure}[hbtp]
  \includegraphics[width=0.5\textwidth]{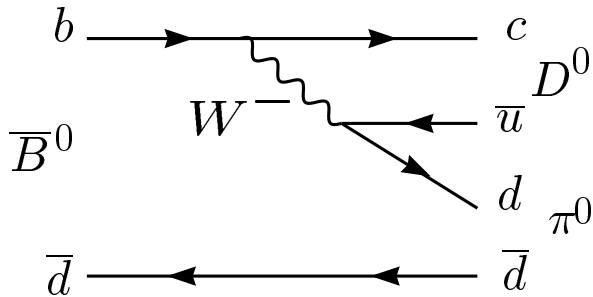}
  \caption{
    \label{diag_fav}
    Diagram for the dominant colour-suppressed amplitude 
    for $\bar{B}^0 \to D\pi^0$.
  }
\end{figure}

Let us now consider the decays of the $B$ meson to $D h^0$.
At first, we consider only the favoured $b\to c\bar{u}d$ (and charge conjugate)
amplitude, shown in Fig.~\ref{diag_fav}.
Then the $D$ meson produced by $B$ decay at time $t_{\rm sig}$
is given by the following admixture:
\begin{equation}
  \label{eq:d_from_b0bar_evo}
  \left| \tilde{D}_{\bar{B}^0}(\dt) \right> = 
  \left| D^0 \right> \cos(\dm\dt/2) - 
  i \frac{p}{q} \eta_{h^0} (-1)^l \left| \bar{D}^0 \right> \sin(\dm\dt/2),
\end{equation}
where we use $\eta_{h^0}$ to denote the $CP$ eigenvalue of $h^0$,
and $l$ gives the orbital angular momentum in the $Dh^0$ system~\cite{dstar}.

The next step is the multibody decay of the $D$ meson.
We use $D \to \kspipi$ for illustration.
We follow~\cite{anton} and describe the amplitude for a 
$\bar{D}^0$ decay to this final state as $f(m_+^2,m_-^2)$, 
where $m_+^2$ and $m_-^2$ are the squares of 
two body invariant masses of the $\ks\pi^+$ and $\ks\pi^-$ combinations.
Assuming no $CP$ violation in the neutral $D$ meson system,
the amplitude for a $D^0$ decay is then given by $f(m_-^2,m_+^2)$.
The amplitude for the $B$ decay at time $t_{\rm sig}$ is then given by
\begin{equation}
  \label{eq:m_b0bar_pre}
  M_{\bar{B}^0}(\dt) = 
  f(m_-^2,m_+^2)  \cos(\dm\dt/2) - 
  i \frac{p}{q} \eta_{h^0} (-1)^l f(m_+^2,m_-^2) \sin(\dm\dt/2).
\end{equation}
Similar expressions for a state which is known to be 
$B^0$ at time $t_{\rm tag}$ 
are obtained by interchanging $B^0 \longleftrightarrow \bar{B}^0$, $D^0 \longleftrightarrow \bar{D}^0$,
$p \longleftrightarrow q$ and $m_+^2 \longleftrightarrow m_-^2$:
\begin{eqnarray}
  \label{eq:b0_evo}
  \left| B^0(\dt) \right> & = &
  e^{-\left|\dt\right|/2\tau_{B^0}}
  \left(
    \left| B^0 \right> \cos(\dm\dt/2) - 
    i \frac{q}{p} \left| \bar{B}^0 \right> \sin(\dm\dt/2)
  \right), \\
  \label{eq:d_from_b0_evo}
  \left| \tilde{D}_{B^0}(\dt) \right> & = &
  \left(
    \left| \bar{D}^0 \right> \cos(\dm\dt/2) - 
    i \frac{q}{p} \eta_{h^0} (-1)^l \left| D^0 \right> \sin(\dm\dt/2)
  \right), \\
  \label{eq:m_b0_pre}
  M_{B^0}(\dt) & = &
  f(m_+^2,m_-^2) \cos(\dm\dt/2) - 
  i \frac{q}{p} \eta_{h^0} (-1)^l f(m_-^2,m_+^2) \sin(\dm\dt/2).
\end{eqnarray}

In the Standard Model, $\left| q/p \right| = 1$ to a good approximation,
and, in the usual phase convention, ${\rm arg}\left(q/p\right) = 2 \phi_1$.
Then
\begin{eqnarray}
  \label{eq:m_b0bar}
  M_{\bar{B}^0}(\dt) & = &
  f(m_-^2,m_+^2)  \cos(\dm\dt/2) - 
  i e^{-i2\phi_1} \eta_{h^0} (-1)^l f(m_+^2,m_-^2) \sin(\dm\dt/2), \\
  \label{eq:m_b0}
  M_{B^0}(\dt) & = &
  f(m_+^2,m_-^2) \cos(\dm\dt/2) - 
  i e^{+i2\phi_1} \eta_{h^0} (-1)^l f(m_-^2,m_+^2) \sin(\dm\dt/2),
\end{eqnarray}
and it can be seen that once the model $f(m_+^2,m_-^2)$ is fixed,
the phase $2\phi_1$ can be extracted from 
a time-dependent Dalitz plot fit to $B^0$ and $\bar{B}^0$ data.

At this point it is instructive to compare to the 
$B^\pm \to DK^\pm$ analysis~\cite{anton}.
In that case we obtained time-independent expressions
\begin{eqnarray}
  \label{eq:m_bminus}
  M_{B^-} & = & f(m_-^2,m_+^2) + r_{DK} e^{i(\delta_{DK} - \phi_3)} f(m_+^2,m_-^2), \\
  \label{eq:m_bplus}
  M_{B^+} & = & f(m_+^2,m_-^2) + r_{DK} e^{i(\delta_{DK} + \phi_3)} f(m_-^2,m_+^2),
\end{eqnarray}
where $r_{DK}$ is the ratio of the magnitudes of 
the contributing (suppressed and favoured) decay amplitudes 
($r_{DK} = \left| A(B^- \to \bar{D}^0 K^-) / A(B^- \to D^0 K^-) \right|$),
and $\delta_{DK}$ is the strong phase between them.
It can be seen that the role of $r_{DK}$
in the time-independent analysis
is taken by the expression $\tan(\dm\dt/2)$ in the time-dependent case.
Furthermore, in the time-dependent case, 
there is no non-trivial strong phase difference.
Therefore, the time-dependent analysis has the advantage
that there is only one unknown parameter,
which partly compensates for the experimental disadvantages that are accrued.

\begin{figure}[hbtp]
  \includegraphics[width=0.5\textwidth]{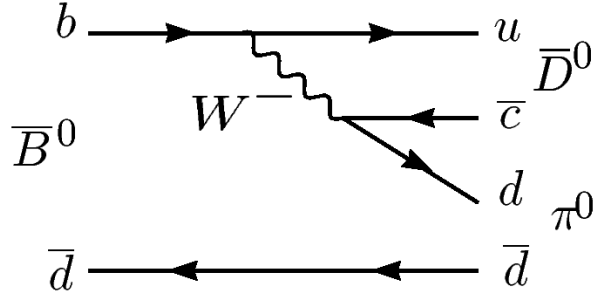}
  \caption{
    \label{diag_sup}
    Diagram for the colour- and Cabibbo-suppressed amplitude 
    for $\bar{B}^0 \to D\pi^0$.
  }
\end{figure}

We now consider the effect of the Cabibbo-suppressed 
$b \to u\bar{c}d$ amplitudes, shown in Fig.~\ref{diag_sup}.
The magnitude of this amplitude is expected to be smaller
than the Cabibbo-favoured diagram (Fig.~\ref{diag_fav})
by a factor of 
\begin{equation}
  r_{D\pi^0} = 
  \frac{
    \left| A(\bar{B}^0 \to \bar{D}^0\pi^0) \right|
  }{
    \left| A(\bar{B}^0 \to D^0\pi^0) \right|
  } \approx \left| \frac{V_{ub}V_{cd}^*}{V_{cb}V_{ud}^*} \right| \approx 0.02.
\end{equation}
Since this simple approximation neglects hadronic factors, 
it is the same for all $h^0$, 
though the precise values will depend on the final state.
We denote the strong phase difference between the two amplitudes as
$\delta_{D\pi^0}$ (which, in general, will be different for each $h^0$).
Including this amplitude, 
the expressions Eqs.~\ref{eq:m_b0bar} and~\ref{eq:m_b0} are replaced by
\begin{eqnarray}
  \label{eq:m_b0bar_full}
  M_{\bar{B}^0}(\dt) & = &
  \left[
    f(m_-^2,m_+^2) + r_{D\pi^0} e^{i(\delta_{D\pi^0} - \phi_3)} f(m_+^2,m_-^2)    
  \right]
  \cos(\dm\dt/2) - \\
  & &
  i e^{-i2\phi_1} \eta_{h^0} (-1)^l 
  \left[
    f(m_+^2,m_-^2) + r_{D\pi^0} e^{i(\delta_{D\pi^0} + \phi_3)} f(m_+^2,m_-^2)
  \right] 
  \sin(\dm\dt/2), \nonumber \\
  \label{eq:m_b0_full}
  M_{B^0}(\dt) & = &
  \left[
    f(m_+^2,m_-^2) + r_{D\pi^0} e^{i(\delta_{D\pi^0} + \phi_3)} f(m_+^2,m_-^2)
  \right] 
  \cos(\dm\dt/2) - \\
  & & 
  i e^{+i2\phi_1} \eta_{h^0} (-1)^l  
  \left[
    f(m_-^2,m_+^2) + r_{D\pi^0} e^{i(\delta_{D\pi^0} - \phi_3)} f(m_+^2,m_-^2)
  \right]
  \sin(\dm\dt/2). \nonumber 
\end{eqnarray}

In principle, therefore, it is possible to extract all four
unknown parameters ($2\phi_1, r_{D\pi^0}, \delta_{D\pi^0}, \phi_3$)
from the time-dependence of the Dalitz plot.
However, due to the smallness of $r_{D\pi^0}$, this is highly impractical.
On the other hand, the above formulation allows us to 
generate simulated data including the suppressed contribution,
and thus estimate the effect of its neglect.

Again, we note that in the case of $\bar{B}^0 \to D\ks$,
the ratio of amplitudes is not small ($r_{D\ks} \sim 0.4$),
and in this case both $\phi_1$ and $\phi_3$ can be extracted from
a time-dependent Dalitz plot analysis.
In fact, the size of $r_{D\ks}$ makes this mode quite attractive
for the measurement of $\phi_3$.
Similarly, a time-independent analysis can be performed 
using $\bar{B}^0 \to D\bar{K}^{*0}$, with $\bar{K}^{*0} \to K^-\pi^+$, 
but in this case additional uncertainty arises due to possible contributions
from nonresonant $\bar{B}^0 \to DK^-\pi^+$~\cite{anton_dkstar}.

\section{Feasibility Study}

The potential accuracy of the $\phi_1$ determination is 
estimated using a Monte Carlo based feasibility study.
We generate $\bdbkspipi$ decays and process the events
with detector simulation and reconstruction.
Signal $B$ candidates are selected.  
Signal and tagging $B$ vertexes are reconstructed in order to obtain $\dt$,
and the flavour of the tagging $B$ meson is obtained. 
Finally, we perform an unbinned likelihood fit of the time-dependent 
Dalitz plot to obtain the value of $\phi_1$ and its uncertainty.

\subsection{Monte Carlo Generation}

\begin{figure}[hbtp]
  \includegraphics[width=0.32\textwidth]{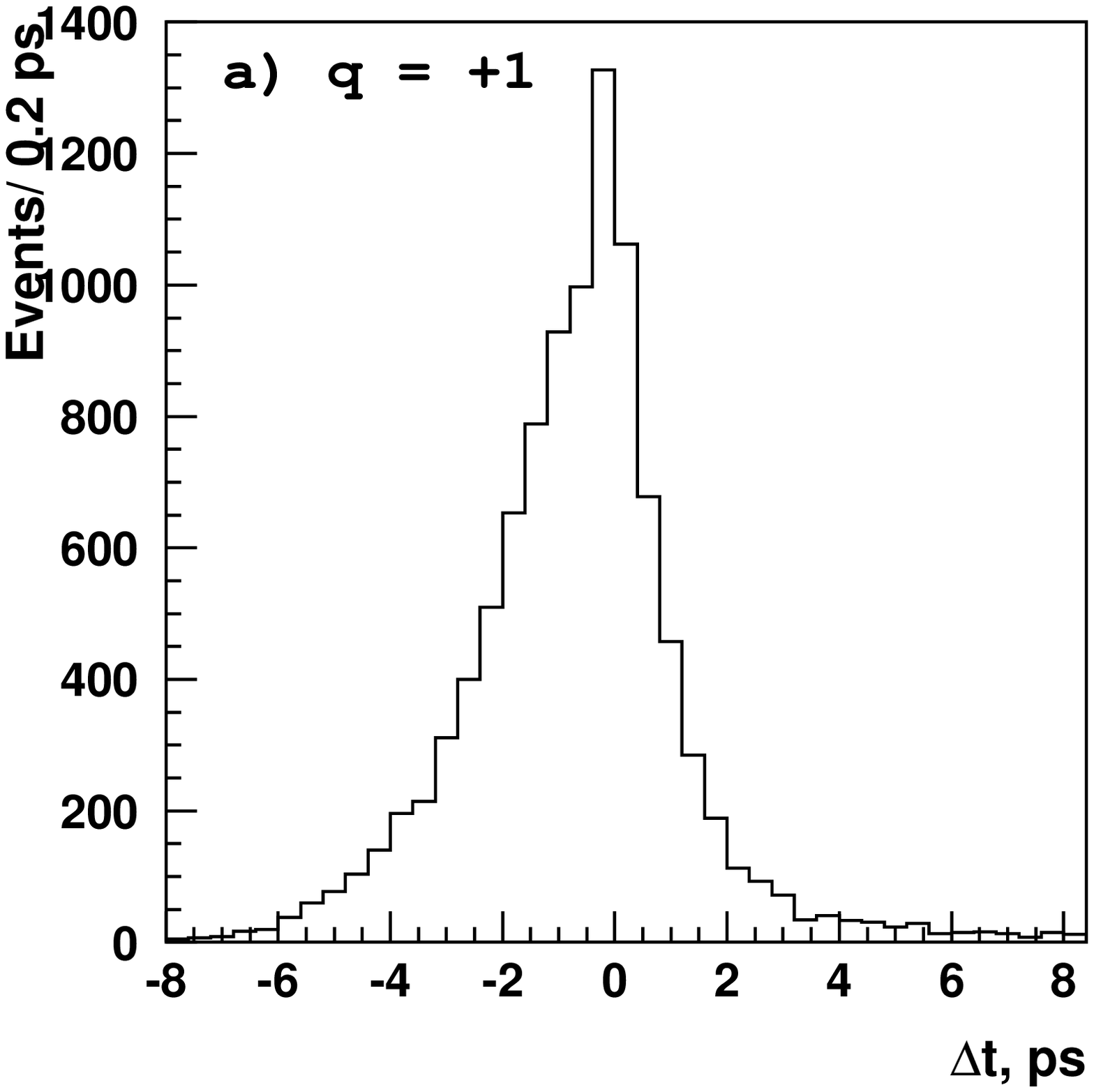}
  \includegraphics[width=0.32\textwidth]{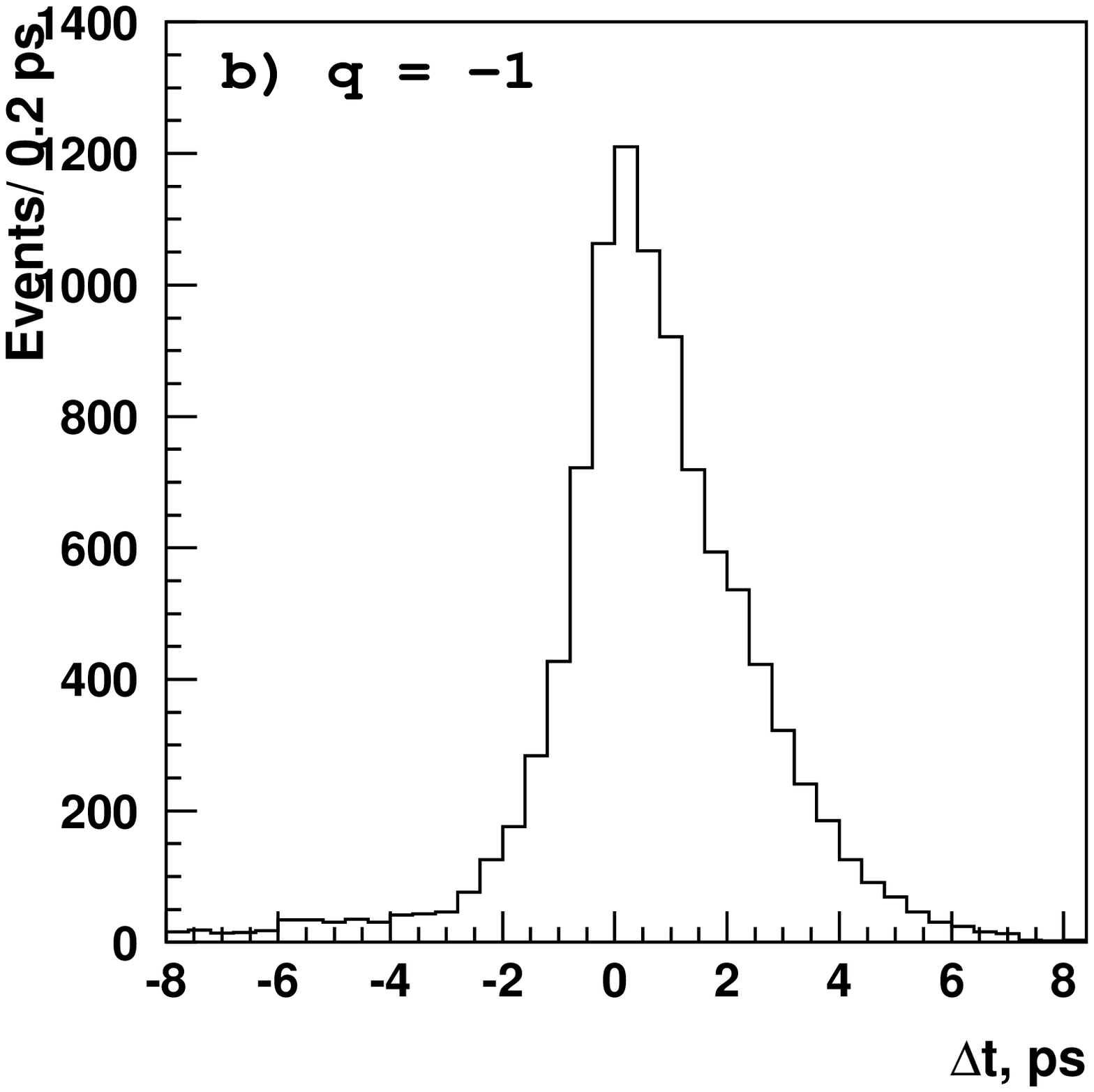}
  \includegraphics[width=0.32\textwidth]{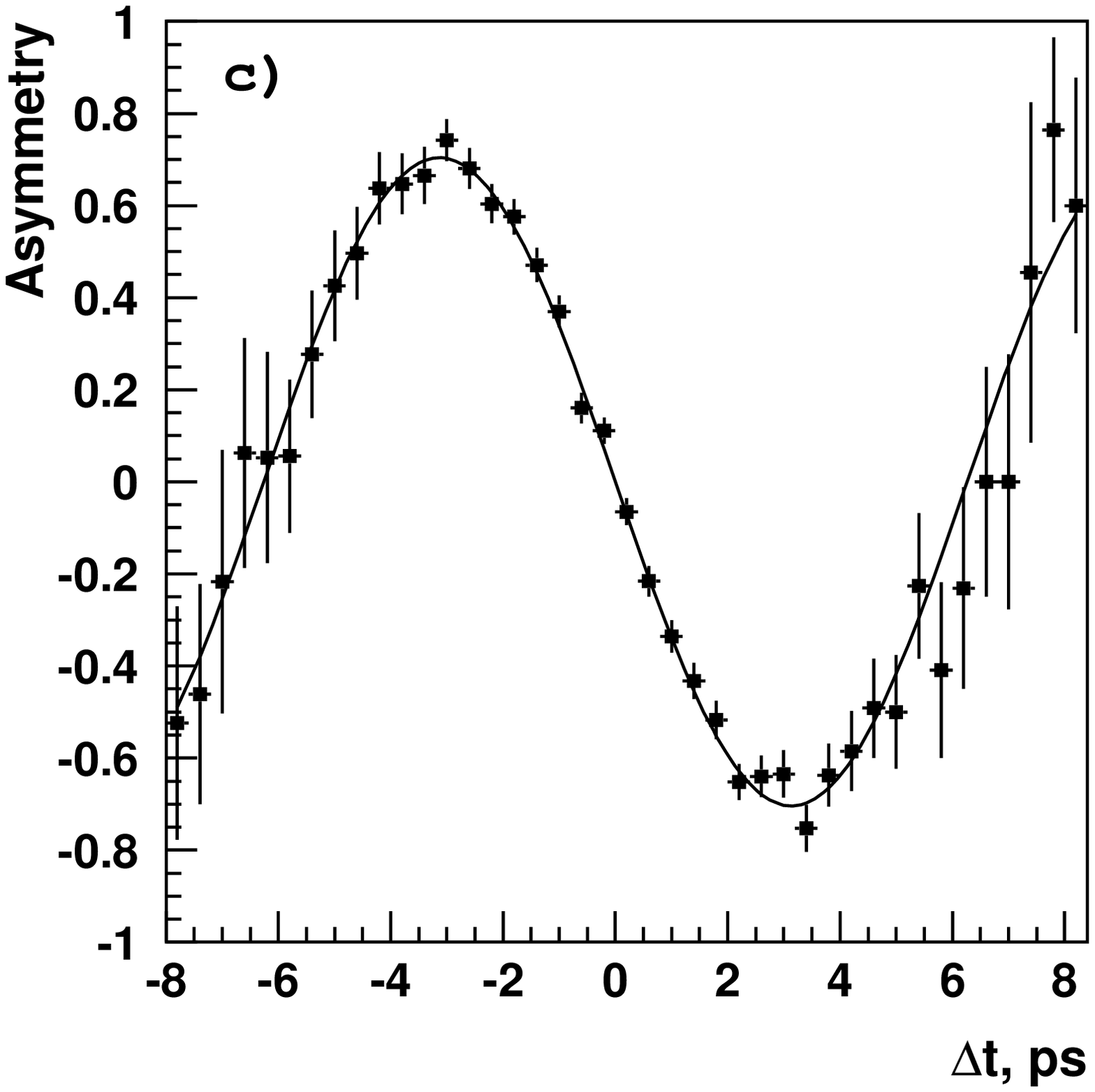}
  \caption{
    \label{dt_ksrho}
    $\dt$ distributions for $\bdbkspipi$
    (a) $\fq = +1$, (b) $\fq = -1$, (c) asymmetry.
  }
\end{figure}

In order to test the feasibility of the method described above,
we have developed an algorithm to generate Monte Carlo simulated data, 
based on {\tt EvtGen}~\cite{evtgen}.
We first test the generator by restricting the $D \to \kspipi$ decay
to the $\ks \rho^0$ channel.
In this case, the formalism simplifies to the familiar $D_{CP}h^0$ case,
and the time-dependent decay rate (neglecting suppressed amplitudes),
is given by
\begin{equation}
  {\cal P}(\dt) = 
  \frac{e^{-\left|\dt\right|/ \tau_{B^0}}}{4\tau_{B^0}}
  \left\{ 1 + q \, {\cal S}_{D_{CP}h^0} \, \sin(\dm\dt) \right\},
\end{equation}
where the $b$-flavour charge $q$ is $+1$ ($-1$) 
when the tagging $B$ meson is $B^0$ ($\bar{B}^0$),
and, within the Standard Model, 
${\cal S}_{D_{CP}h^0} = - \eta_{D_{CP}} \eta_{h^0} (-1)^l \sin(2\phi_1)$.
For the $CP$ odd decay $D \to \ks\rho^0$, $\eta_{D_{CP}} = -1$,
so for $\left( \ks\rho^0 \right)_D \pi^0$, ${\cal S}_{D_{CP}h^0} = - \sin(2\phi_1)$.
Fig.~\ref{dt_ksrho} shows generator level information for these decays.

\begin{figure}[hbtp]
  \includegraphics[width=0.35\textwidth]{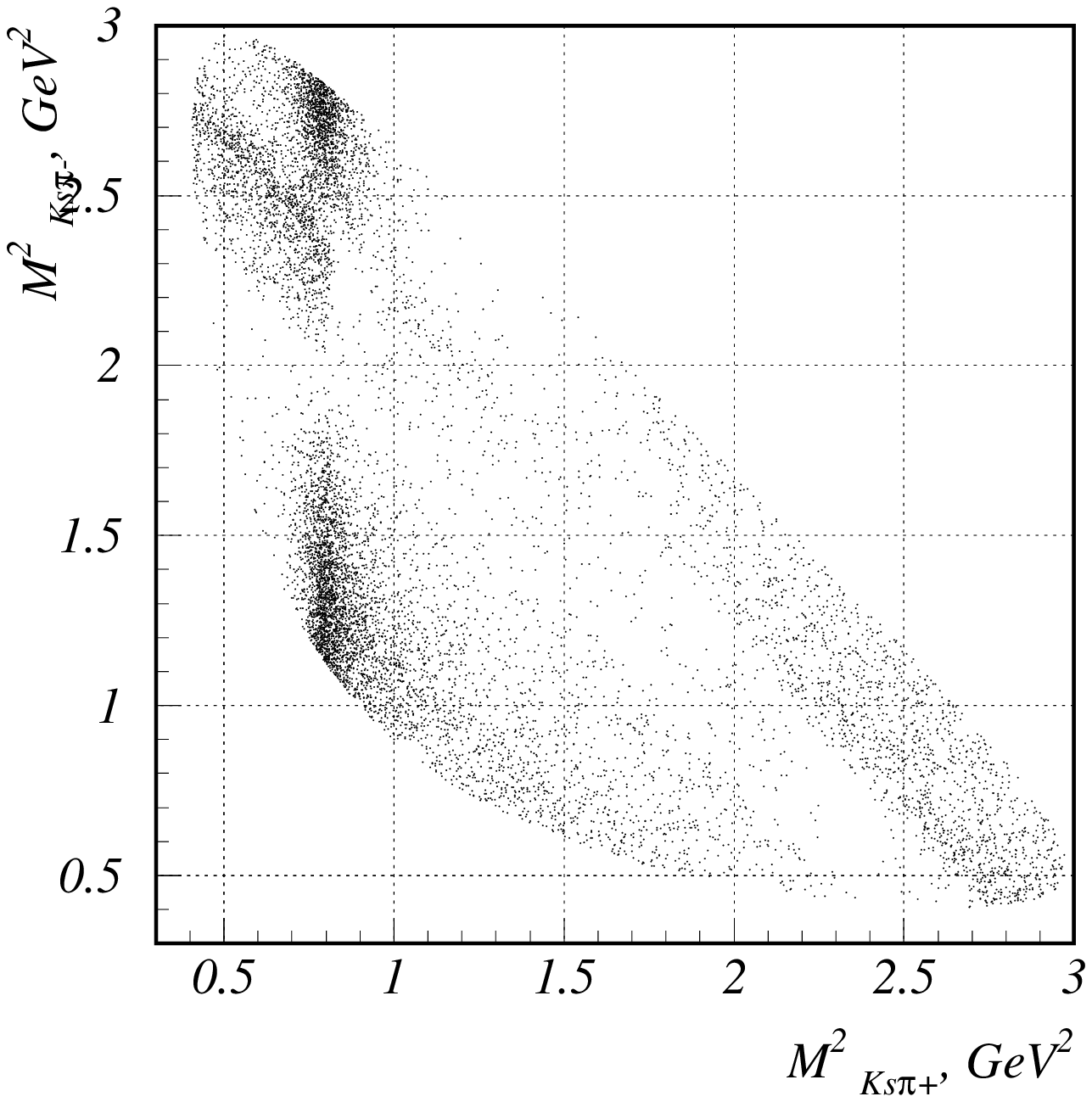}
  \caption{
    \label{dmodel}
    $\dbkspipi$ decay Dalitz plot.
  }
\end{figure}

We next implement three body $D$ decays into our generator.
The amplitude of the $\dbkspipi$ decay is described by a coherent 
sum of two-body decay amplitudes plus non-resonant part:
\begin{equation}
  f(m^2_{\ks\pi^+}, m^2_{\ks\pi^+})=
  \sum\limits_{j=1}^{N} a_j e^{i\alpha_j}A_j
  (m^2_{\ks\pi^+}, m^2_{\ks\pi^+})+b e^{i\beta}, 
\end{equation}
where $N$ is the number of resonances, 
$A_j(m^2_{\ks\pi^+}, m^2_{\ks\pi^+})$, $a_j$ and 
$\alpha_j$ are the matrix element, amplitude and phase, respectively, 
for the $j$-th resonance, and $b$ and $\beta$ are the amplitude
and phase for the non-resonant component. 
For further details, see~\cite{anton} and references therein.
Table~\ref{reslist} describes the set of resonances we use in 
the decay model of our generator,
which is similar to that in the CLEO measurement~\cite{dkpp_cleo}.
Fig.~\ref{dmodel} shows the Dalitz plot distribution for the
$\dbkspipi$ decay generated according this model.

\begin{table}[hbtp]
  \begin{center}
    \caption{
      \label{reslist}
      List of resonances used for $\dbkspipi$ decay simulation.
    }
    \vspace{0.5\baselineskip}
    \begin{tabular}{|l|c|c|} \hline
      Resonance                    & Amplitude & Phase ($^\circ$) \\ \hline
      $K^*(892)^+\pi^-$            & 1.418     & 170       \\ 
      $K_0^*(1430)^+\pi^-$         & 1.818     &  23       \\ 
      $K_2^*(1430)^+\pi^-$         & 0.909     & 194       \\ 
      $K^*(892)^-\pi^+$ (DCS)      & 0.100     & 341       \\
      $\ks\rho^0$                  & 0.909     &  20       \\
      $\ks\omega$                  & 0.034     & 134       \\
      $\ks f_0(980)$               & 0.309     & 208       \\
      $\ks f_0(1370)$              & 1.636     & 105       \\
      $\ks f_2(1270)$              & 0.636     & 328       \\
      $\ks\pi^+\pi^-$ non-resonant & 1.0       &   0       \\ \hline
    \end{tabular}
  \end{center}
\end{table}

\begin{figure}[hbtp]
  \includegraphics[width=0.32\textwidth]{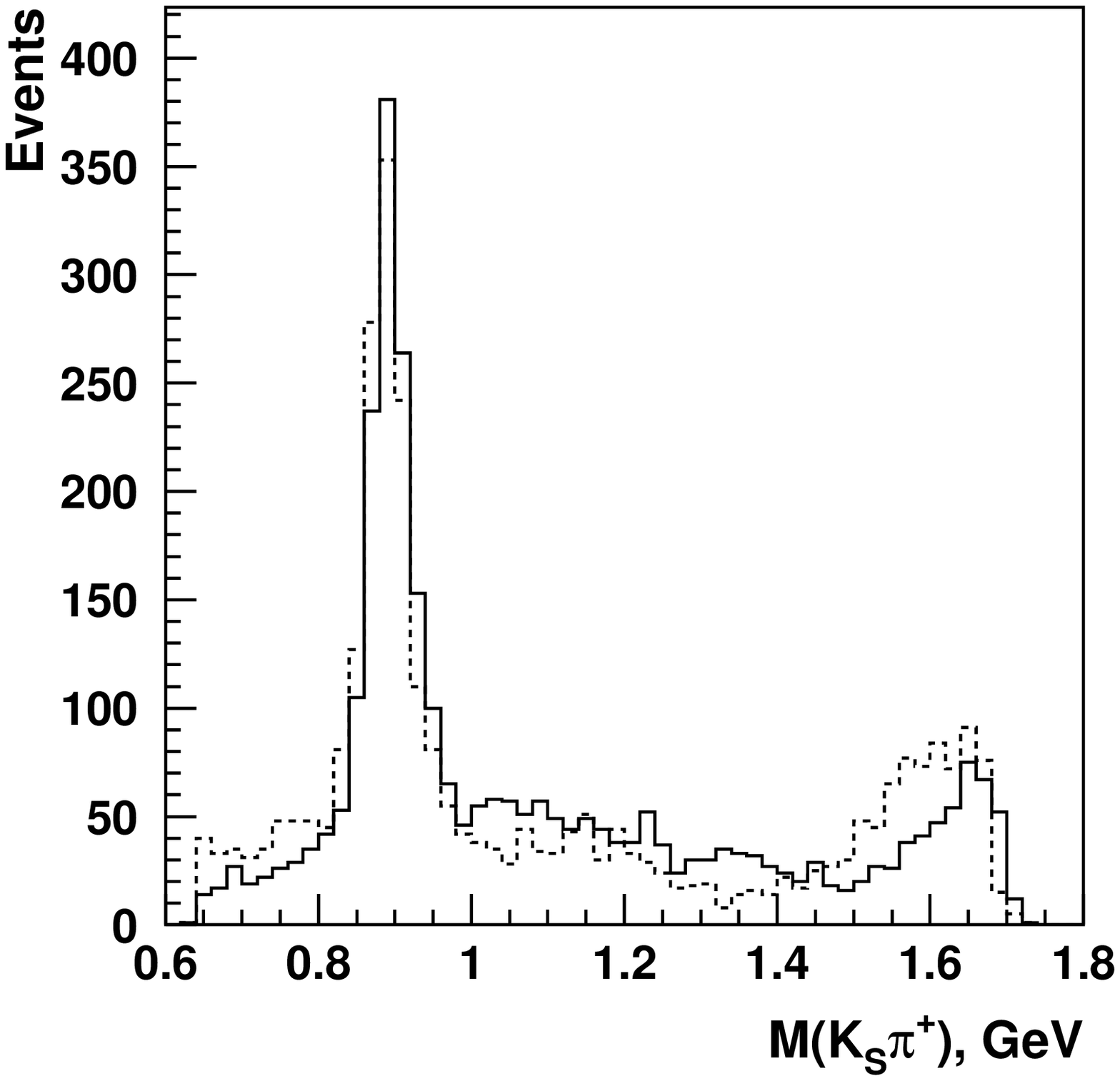}
  \includegraphics[width=0.32\textwidth]{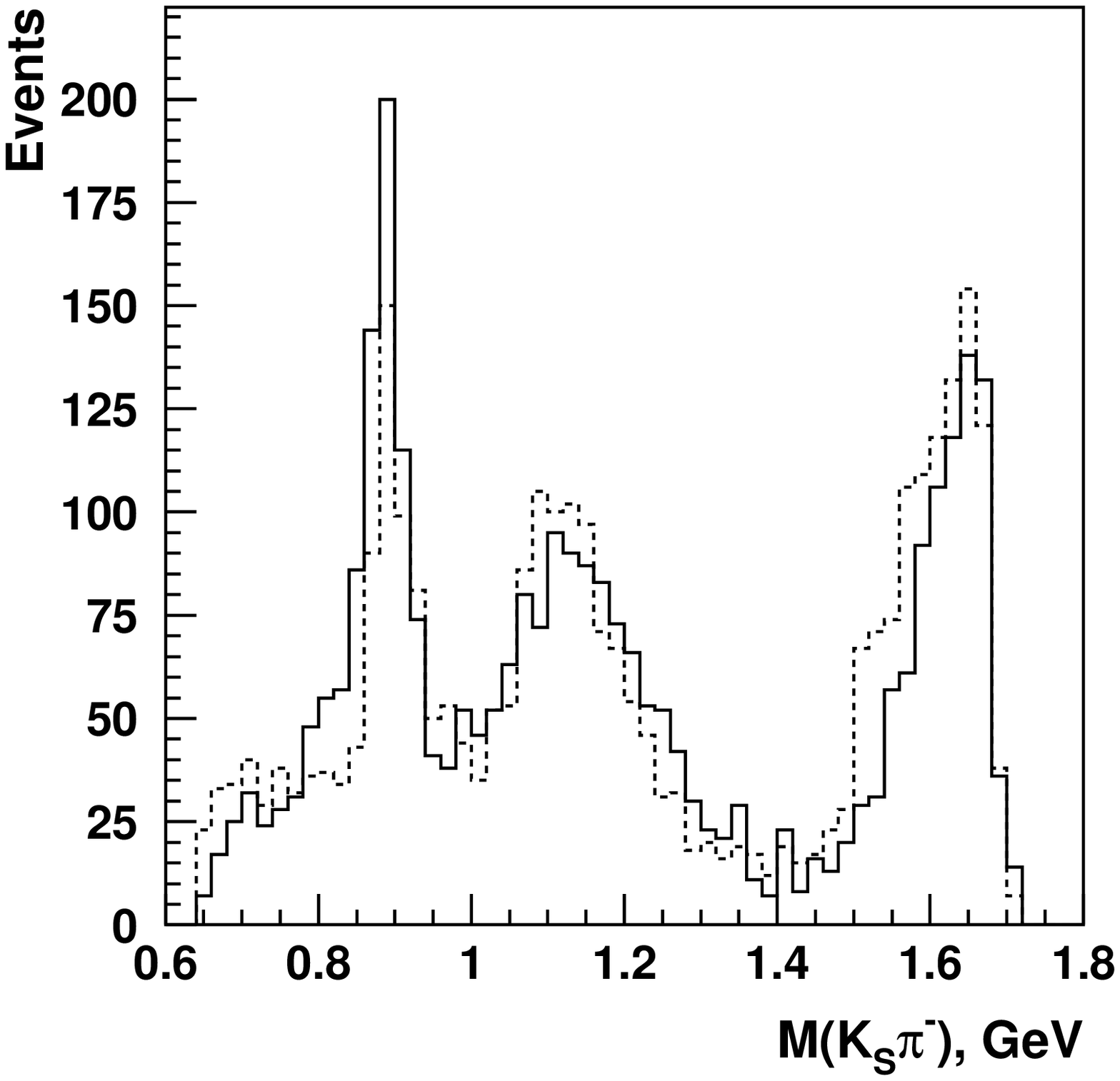}
  \includegraphics[width=0.32\textwidth]{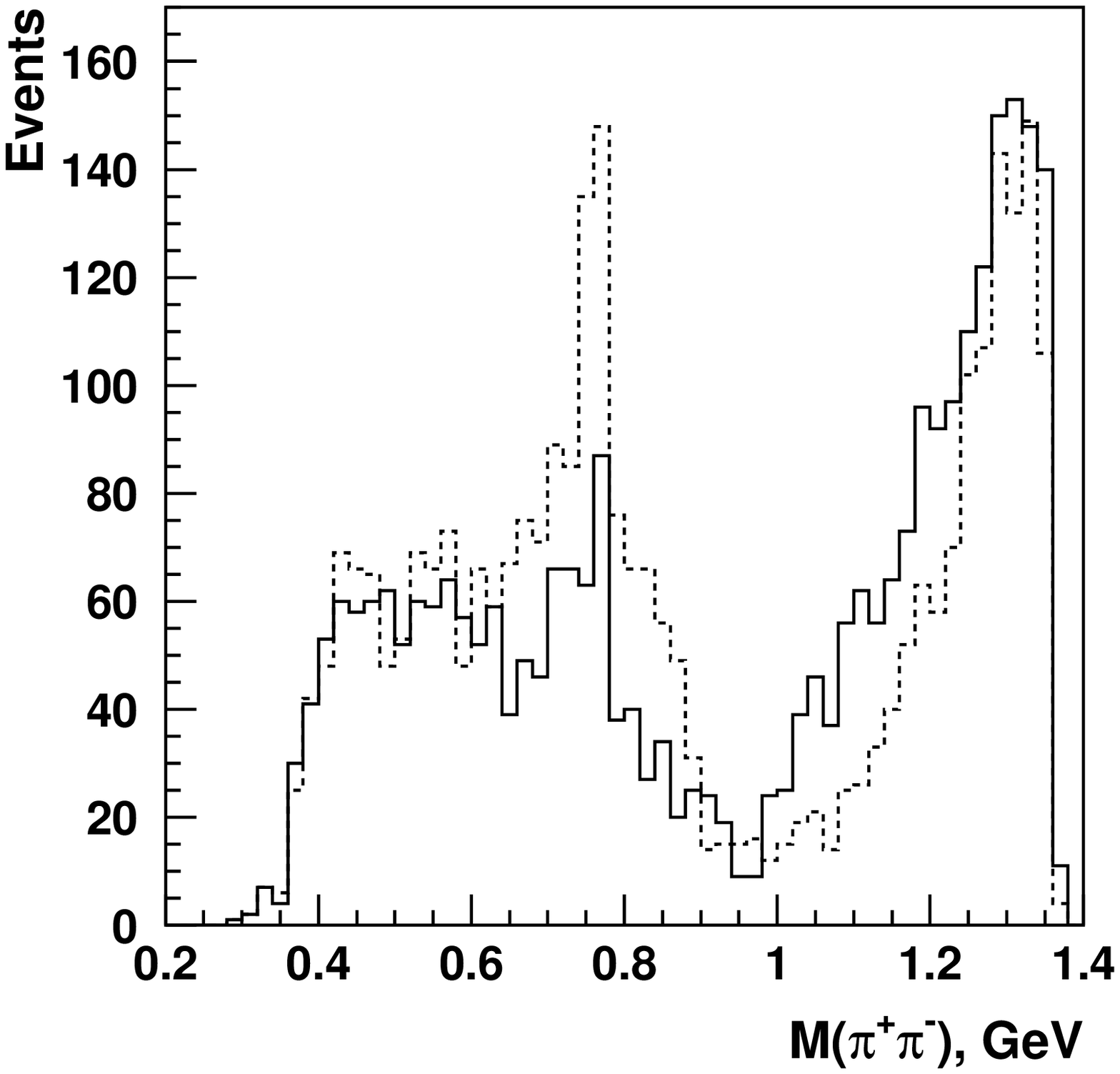}
  \caption{
    \label{time_dalitz}
    Generator level invariant mass distributions of $D$ decay daughters
    produced in the $\bdbkspipi$ decay chain.
    Events are generated with $2\phi_1=47^o$, and only 
    events with $q = -1$ (tagging $B$ decays as $\bar{B}^0$) are shown. 
    The dashed histograms show distributions for events with $\dt>\tau_{B^0}/2$, 
    the solid histograms show those for events with $\dt<-\tau_{B^0}/2$.
  }
\end{figure}

For further confirmation of the operation of our generator,
we look at the generator level time-dependent Dalitz plot.
We generate $\bdbkspipi$ decays using $2\phi_1 = 47^\circ$.
In Fig.~\ref{time_dalitz} we show the invariant mass distributions
of the $D$ decay daughters for events with $q = -1$,
and compare those for events with $\dt$ greater than $\tau_{B^0}/2$ 
with those for events with $\dt$ less than $-\tau_{B^0}/2$. 
Events with $\left| \dt \right| < \tau_{B^0}/2$ or $q = +1$ are not shown.
We see clear differences in the two invariant mass distributions;
in particular we see more events with positive than negative $\dt$
in the $\rho^0$ region of the $\pi^+\pi^-$ invariant mass distribution,
as expected from Fig.~\ref{dt_ksrho}.

Since we are concerned with the feasibility of studying these modes
at $B$ factory experiments, we use the software of the Belle collaboration
to perform simulation of the Belle detector
and to reconstruct candidate events.
The Belle detector is a large-solid-angle magnetic spectrometer that
consists of a silicon vertex detector (SVD), 
a central drift chamber (CDC),
aerogel threshold \v{C}erenkov counters (ACC),
time-of-flight scintillation counters (TOF),
and an electromagnetic calorimeter (ECL)
located inside a superconducting solenoid coil that provides a
$1.5 \ {\rm T}$ magnetic field.
An iron flux-return located outside of the coil is instrumented
to detect $K_L^0$ mesons and to identify muons (KLM).
The detector is described in detail elsewhere~\cite{belle}.
The detector simulation is based on GEANT~\cite{geant}.
Belle is installed at the interaction point of the KEKB asymmetric-energy
$e^+e^-$ ($3.5 \ {\rm GeV}$ on $8 \ {\rm GeV}$) collider~\cite{KEKB}.
KEKB operates at the $\Upsilon(4S)$ resonance
($\sqrt{s} = 10.58 \ {\rm GeV}$) with a peak luminosity that exceeds
$1.5 \times 10^{34}~{\rm cm}^{-2}{\rm s}^{-1}$.
The asymmetric energy allows $\dt$ to be determined from the 
displacement between the signal and tagging $B$ meson decay vertices.

\subsection{Event Reconstruction}

We reconstruct the decays $\bdbkspipi$ for $h^0 = \pi^0, \eta$ and $\omega$.
We use the subdecays $\ks \to \pi^+\pi^-$, $\pi^0 \to \gamma\gamma$, $\eta \to \gamma\gamma, \pi^+\pi^-\pi^0$ 
and $\omega \to \pi^+\pi^-\pi^0$.
The reconstruction, including suppression of the dominant background 
from $e^+e^- \to \qq$ ($q = u,d,s,c$) continuum processes, 
is highly similar to that in related Belle analyses~\cite{belle_dh0}.
The properties of the background events are studied using 
generic $B\bar{B}$ and $q\bar{q}$ Monte Carlo.
Our studies allow us to estimate the number of signal and background
events to expect from a given data sample
(we use the data sample of $253 \ {\rm fb}^{-1}$,
containing 275 million $B\bar{B}$ pairs,
collected with the Belle detector before summer 2004 as our baseline).
The results are summarised in Table~\ref{tabeff}.

\begin{table}[hbtp]
  \caption{
    \label{tabeff}
    Detection efficiency, expected numbers of signal ($N_{\rm sig}$) 
    and background ($N_{\rm bkg}$) events 
    and signal purity for the $\bdbkspipi$ final states.
    The expected numbers of events are based on the Belle data sample
    of $253 \ {\rm fb}^{-1}$.
  }
  \vspace{0.5\baselineskip}
  \begin{tabular}
    {|l|c|@{\hspace{3mm}}c@{\hspace{3mm}}|@{\hspace{3mm}}c@{\hspace{3mm}}|c|}
    \hline
    Process & Efficiency (\%) & $N_{\rm sig}$ & $N_{\rm bkg}$ & Purity \\ 
    \hline
    $D\pi^0$ & 8.1 & 118 & 49 & 71\% \\
    $D\omega$   & 3.9 &  49 &  8 & 86\% \\
    $D\eta$   & 4.3 &  47 & 15 & 76\% \\
    \hline
    Sum    &     & 214 & 72 & 75\% \\
    \hline
  \end{tabular}
\end{table}

For our further studies, we use only the $D\pi^0$ mode,
for which the expected number of signal events is the largest.
In our pseudo-experiments, described below,
we use numbers of signal and background events (300 and 100 respectively)
which are rounded up from the totals in Table~\ref{tabeff},
as we expect some improvement is possible due to 
optimization of the selection for this analysis.

The signal $B$ meson decay vertex is reconstructed using the $D$
trajectory and an interaction profile (IP) constraint. 
The tagging $B$ vertex position is obtained with the IP constraint and 
with well reconstructed tracks that are not assigned to signal $B$ candidate.
The algorithm is described in detail elsewhere~\cite{resol}.

Tracks that are not associated with the reconstructed $\bdbkspipi$ decay
are used to identify the $b$-flavour of the accompanying $B$ meson.
The tagging algorithm is described in detail elsewhere~\cite{flavor}.
We use two parameters, $\fq$ and $r$, to represent the tagging 
information.
The first, $q$, has the discrete value $+1$~($-1$)
when the tagging $B$ meson is more likely to be a $B^0$~($\bar{B}^0$).
The parameter $r$ corresponds to an event-by-event 
flavour-tagging dilution that ranges
from $r=0$ for no flavour discrimination
to $r=1$ for an unambiguous flavour assignment.

We divide the $\phi_1=[0^\circ:180^\circ]$ range into 18 points in steps of $10^\circ$. 
For each point we perform 30 pseudo-experiments with data samples
consisting of 300 reconstructed $D\pi^0$ events. 
We add 100 background events to each sample,
where the background is modelled by $\bdbpn$, 
with uniform phase space decay $\db\to\kspipi$.

For each pseudo-experiment,
we perform a unbinned time-dependent Dalitz plot fit.
The inverse logarithm of the unbinned likelihood function is minimized:
\begin{equation}
  -2 \log L = 
  -2 
  \left[ 
    \sum \limits^n_{i=1} \log p(m^2_{+i}, m^2_{-i}, \dt_i) - 
    \log \int \limits_D p(m^2_+, m^2_-, \dt) dm^2_+ dm^2_- d\dt 
  \right], 
\end{equation}
where $n$ is the number of events,
$m^2_{+i}$, $m^2_{-i}$ and $\dt_i$ 
are the measured invariant masses of the $D$ daughters,
and the time difference between signal and tagging $B$ meson decays. 
The function $p(m^2_+, m^2_-, \dt)$ is the time-dependent Dalitz plot density,
which is based on Eqs.~\ref{eq:m_b0bar} and~\ref{eq:m_b0},
including experimental effects such as mistagging and $\dt$ resolution - 
we use the standard Belle algorithms to take these effects into account.
The background component is also introduced into $p$.

\begin{figure}[hbtp]
  \includegraphics[width=0.48\textwidth]{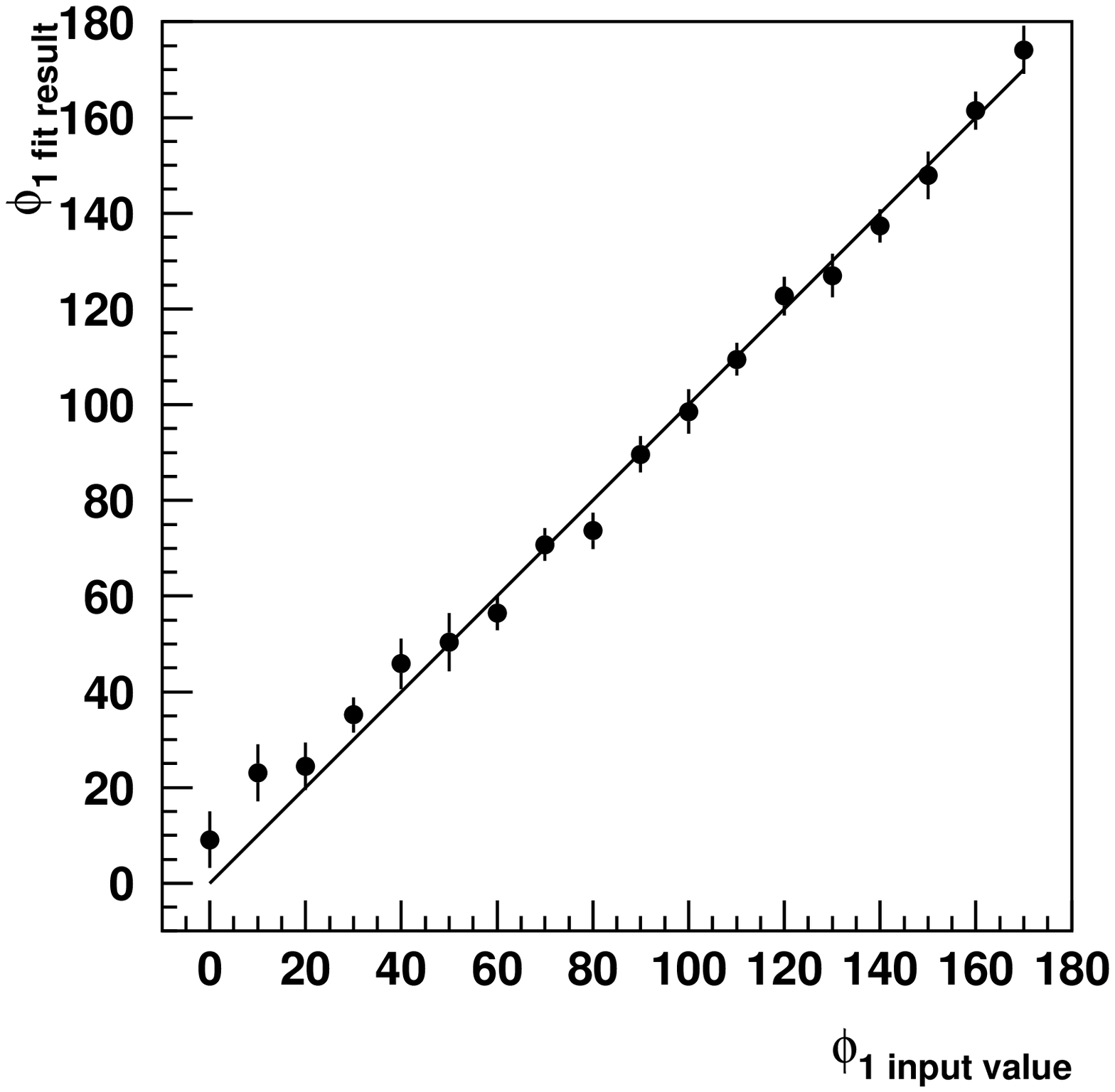}
  \includegraphics[width=0.48\textwidth]{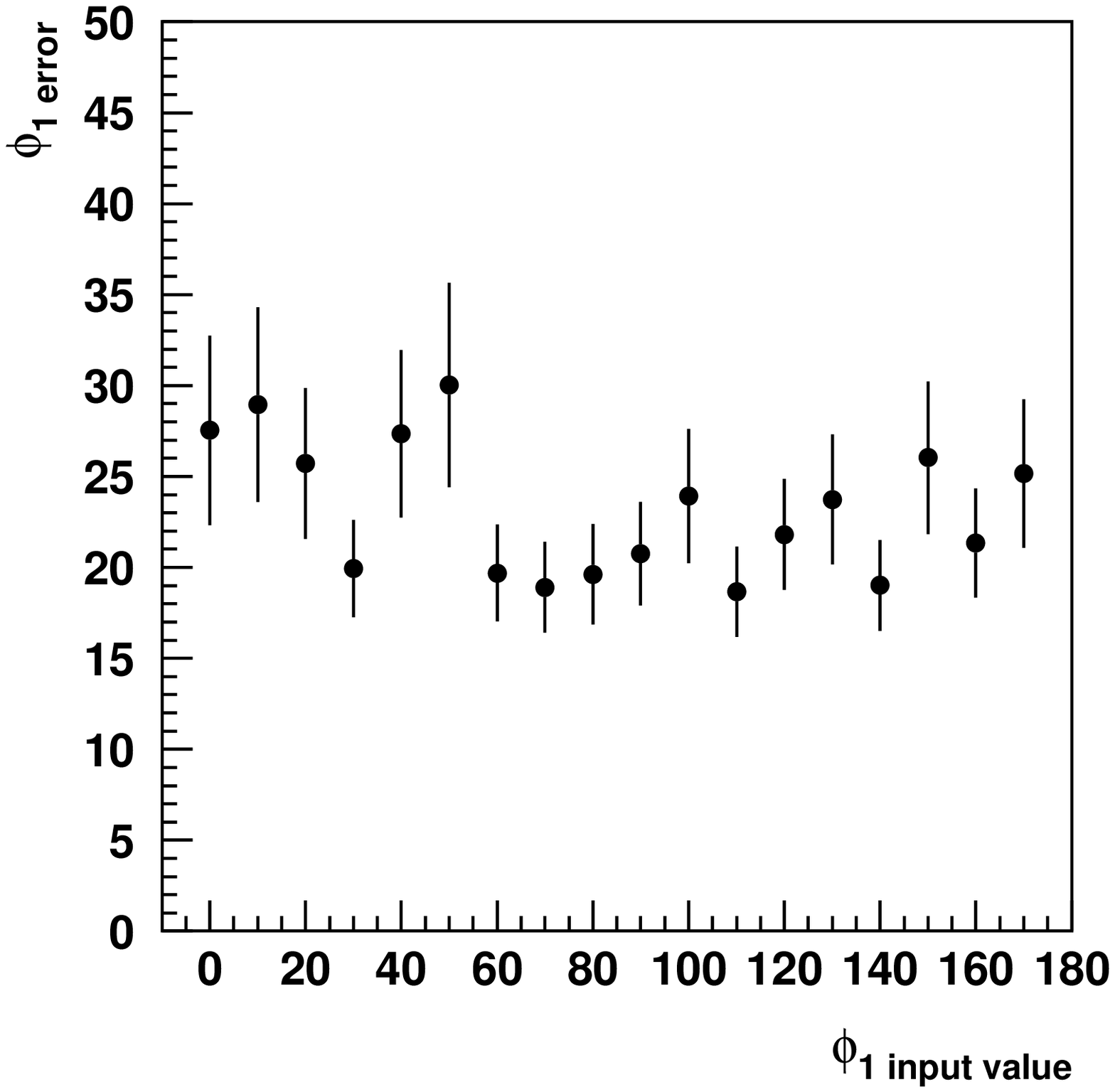}
  \caption{
    \label{phi1res}
    (Left) average $\phi_1$ fit result and (right) $\phi_1$ statistical error,
    as functions of the input value.
  }
\end{figure}

Thus, for each input value of $\phi_1$ 
we obtain fitted results from 30 pseudo-experiments.
From the means and widths of the distributions of these results 
we obtain the average $\phi_1$ fit results and 
estimates of their statistical errors.
These results are shown in Fig.~\ref{phi1res}.
We find the fit results are in good agreement with the input values,
and the expected uncertainty on $\phi_1$ is around $25^\circ$.

\begin{figure}[hbtp]
  \includegraphics[width=0.48\textwidth]{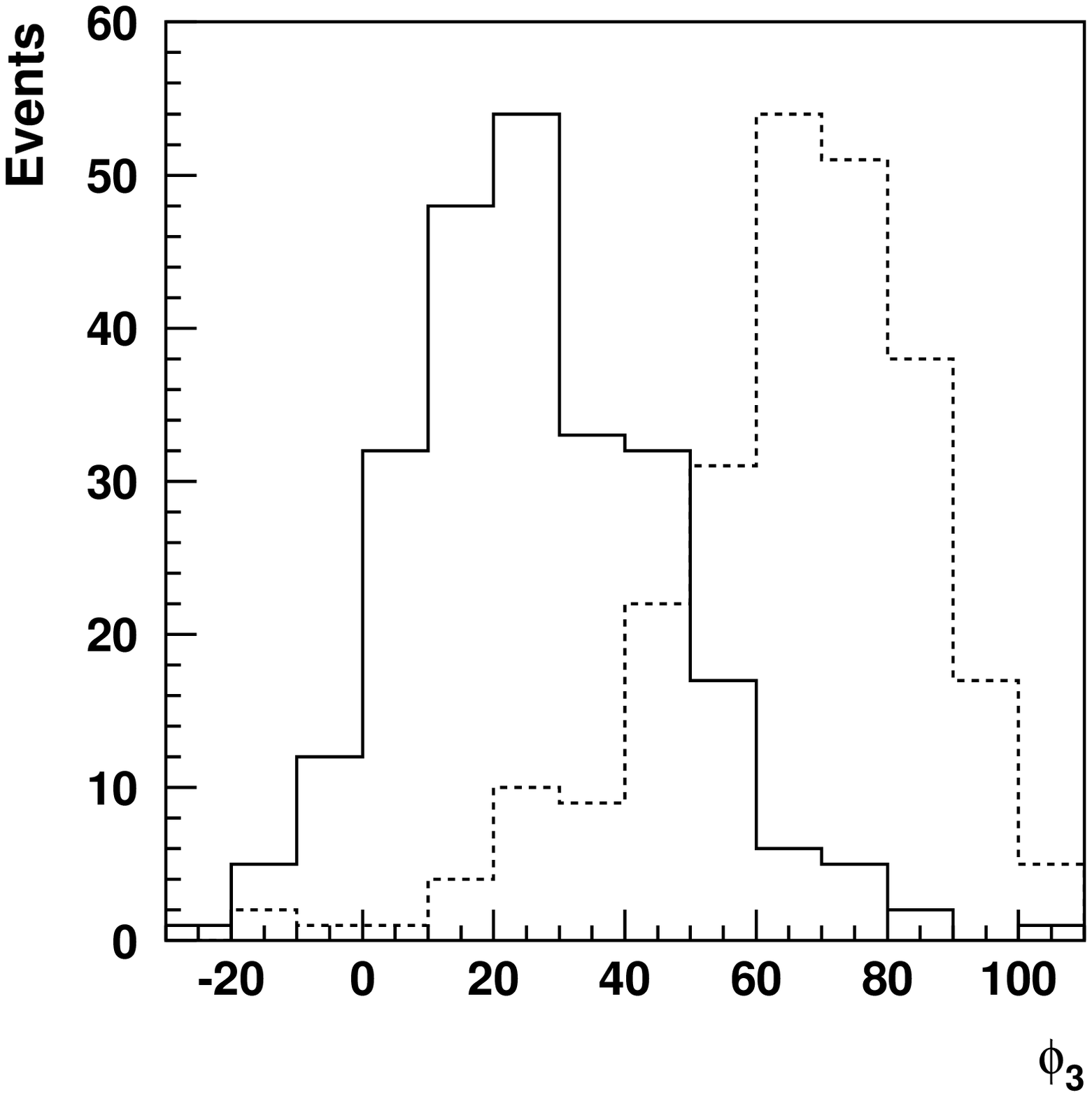}
  \includegraphics[width=0.48\textwidth]{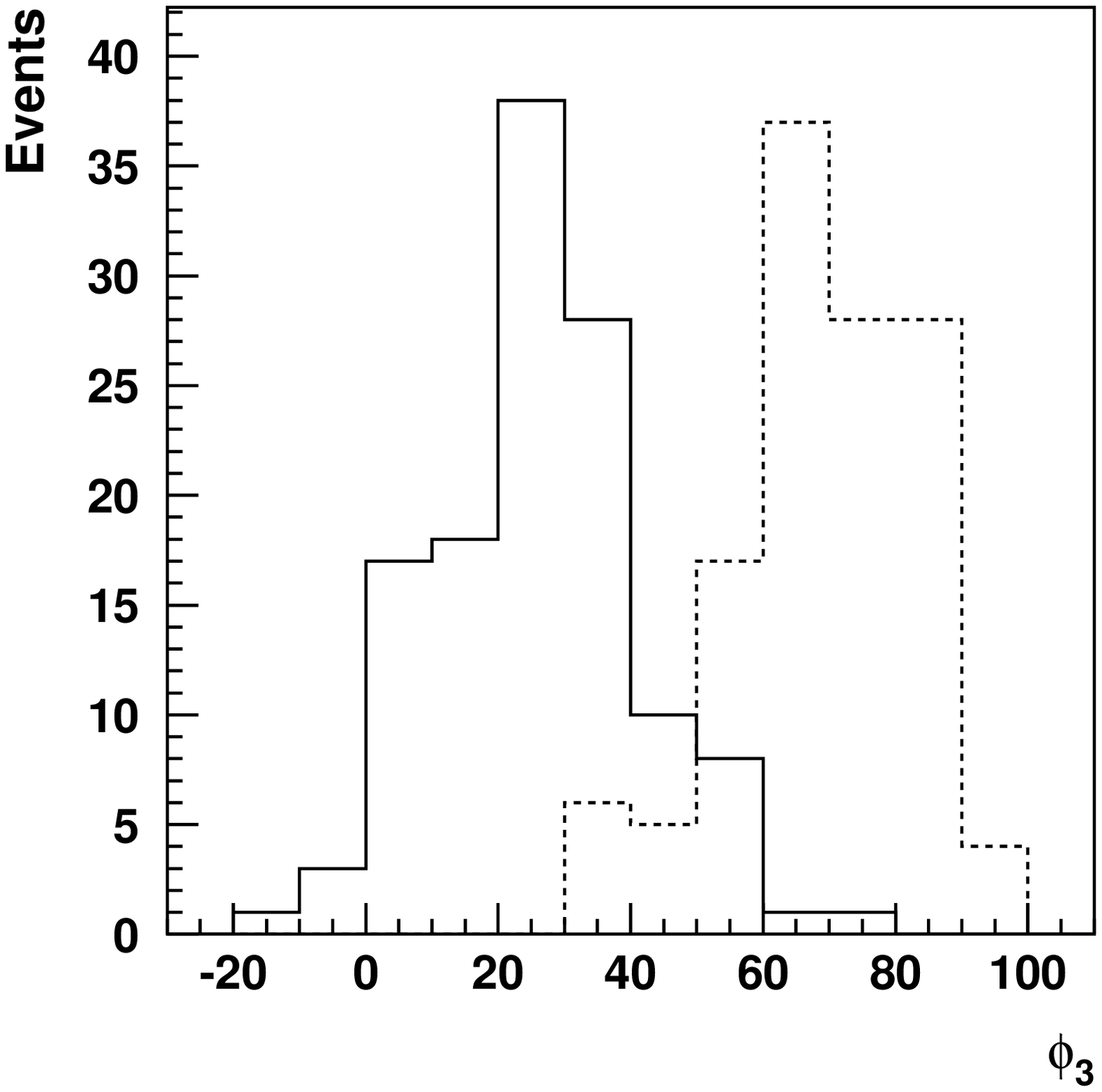}
  \caption{
    \label{pdf}
    Fit results for $\phi_1$.
    The solid (hatched) histograms correspond 
    to the input value $2\phi_1=47^\circ$ ($2\phi_1=133^\circ$).
    The left (right) plot corresponds to a data sample 
    roughly equivalent to $250 \ {\rm fb}^{-1}$ ($500 \ {\rm fb}^{-1}$).
    }
\end{figure}

To look for tails in the distributions,
we also study larger ensembles of pseudo-experiments 
for two $\phi_1$ input values: 
$23.5^\circ$ and $66.5^\circ$, which correspond to $\sin(2\phi_1)=0.73$. 
We have performed this study both for the numbers of events
described above, corresponding roughly to $250 \ {\rm fb}^{-1}$
(for which we perform 250 pseudo-experiments for each input value of $\phi_1$),
and for numbers twice larger (hence $500 \ {\rm fb}^{-1}$,
for which we perform 125 pseudo-experiments).
Fig.~\ref{pdf} shows the distribution of the fit results.
We do not observe any pathological behaviour,
demonstrating that this method can indeed be used to distinguish the two 
solutions for $\sin(2\phi_1)$ with sufficiently large data samples.

We have tested for possible bias in the method due to neglect
of the suppressed amplitudes 
(Eqs.~\ref{eq:m_b0bar_full} and~\ref{eq:m_b0_full}).
Due to the smallness of $r_{D\pi^0}$ compared to the $B^0 - \bar{B}^0$
mixing effect, we expect any such bias to be small,
and indeed we find it to be smaller than $1\%$.

As noted above, this method is highly similar to that used to extract $\phi_3$,
using $B^\pm \to D K^\pm$ followed by multibody $D$ decay~\cite{anton,ggsz}.
A significant complication arises in that case due to uncertainty
in the $D$ decay model, 
and we expect this will also affect the $\bar{B}^0 \to D h^0$ analysis.
However, the time-dependent analysis does not suffer due to 
the smallness of the ratio of amplitudes,
and therefore we expect that the model uncertainty may be smaller.
Furthermore, a number of methods have been proposed to address
the model uncertainty 
(for example, using information from $CP$ tagged $D$ mesons
which can be studied at a $c\tau$ factory, such as CLEO-c), 
and this analysis can also take advantage of any progress in that area.

\section{Conclusion}
We have presented a new method to measure the Unitarity Triangle angle $\phi_1$ 
using amplitude analysis of the multibody decay 
of the neutral $D$ meson produced in the processes $\bar{B}^0 \to Dh^0$. 
The method is directly sensitive to the value of $2\phi_1$ 
and can thus be used to resolve the discrete ambiguity $2\phi_1 \longleftrightarrow \pi-2\phi_1$. 
The expected precision of this method has been studied using 
Monte Carlo simulation.
We expect the uncertainty on $\phi_1$ to be about $25^\circ$ 
for an analysis using a data sample of $253 \ {\rm fb}^{-1}$.

\section*{Acknowledgements}
We are grateful to our colleagues in the Belle Collaboration,
on whose software our feasibility study is based.
We thank the KEK computer group and the 
National Institute of Informatics for valuable computing
and Super-SINET network support. 
We acknowledge support from the 
Ministry of Education, Culture, Sports, Science, and Technology of Japan 
and the Japan Society for the Promotion of Science, and from the 
Ministry of Science and Education of the Russian Federation.
We thank A.~Poluektov for his help with details of the Dalitz analysis,
and T.~Browder and Y.~Sakai for useful suggestions.
We thank Y.~Grossman and A.~Soni for interesting discussions.


\begin{thebibliography}{100}
\bibitem{bs}
  A full treatment of the $B_s$ case must take 
  the non-zero lifetime difference into account.
  We do not include this extension here, for brevity.
\end{thebibliography}
\end{document}